\documentclass[12pt,aps,showpacs,showkeys,preprintnumbers,amsmath,amssymb]{revtex4}

\usepackage{graphicx}

\def\pmb#1{\setbox0=\hbox{$#1$}%
  \kern-.025em\copy0\kern-\wd0
  \kern.05em\copy0\kern-\wd0
  \kern-.025em\raise.0433em\box0}

\def\beq{\begin{equation}}
\def\eeq{\end{equation}}

\begin{document}

\title{Spacetime Splitting, Admissible Coordinates and
Causality}

\author{D. Bini}
\affiliation{
Istituto per le Applicazioni del Calcolo ``M. Picone,'' CNR, I-00185 Rome, Italy\\
ICRA, University of Rome ``La Sapienza,'' I-00185 Rome, Italy\\
INFN, Sezione di Firenze, I--00185 Sesto Fiorentino (FI), Italy}

\author{C. Chicone}
\affiliation{Department of Mathematics and Department of Physics and Astronomy, University of Missouri, Columbia,
Missouri 65211, USA}

\author{B. Mashhoon}
  \affiliation{Department of Physics and Astronomy, University of Missouri, Columbia, Missouri 65211, USA}

\date{\today}

\begin{abstract}
To confront relativity theory with observation, it is necessary to split spacetime into its temporal and spatial components. The (1+3) timelike threading approach involves restrictions on the gravitational potentials $(g_{\mu \nu})$, while the (3+1) spacelike slicing approach involves  restrictions on $(g^{\mu \nu})$. These latter coordinate conditions protect chronology within any such coordinate patch. While the threading coordinate conditions can be naturally integrated into the structure of Lorentzian geometry and constitute the standard coordinate conditions in general relativity, this circumstance does not extend to the slicing coordinate conditions. We explore the influence of chronology violation on wave motion. In particular, we consider the propagation of radiation parallel to the rotation axis of stationary G\"odel-type universes characterized by parameters $\eta > 0$ and $\lambda > 0$ such that for $\eta < 1$ ($\eta >1$) chronology is protected (violated). We show that in the WKB approximation such waves can freely propagate only when chronology is protected. 
\end{abstract}



\maketitle

\section{Introduction}

Though space and time refer to essentially different aspects of our experience, their unification in the theory of relativity has been a distinct achievement. To interpret observations in accordance with this theory, however, we need to split spacetime into its components~\cite{mfg}. Observers are macrophysical entities; therefore, their associated temporal coordinate is thermodynamic time derived from the concept of entropy. This leads to the arrow of time, its one-way property that has no analog in space. In principle, classical gravity can change this one-way character of time and produce a closed timelike curve (``CTC"). That is, such a possibility is not ruled out by the geometric structure of classical general relativity. In this paper, we approach the gravitational violation of chronology from the viewpoint of the theory of measurement of space and time.

We begin our discussion of spacetime splitting with ideal inertial observers in Minkowski spacetime.  Imagine a global inertial frame with coordinates $x^\mu=(ct, \mathbf{x})$ and spacetime interval $-ds^2=\eta_{\mu \nu} dx^\mu dx^\nu$, where $(\eta_{\mu \nu}) = {\rm diag}\, (-1, 1, 1, 1)$. We define \emph{fundamental observers} to be those at rest in space. The world lines of the fundamental observers are therefore temporal coordinate lines that thread spacetime. At each instant of coordinate time $t$, the corresponding orthogonal subspace is the ``space" of the static observer. This is schematically depicted in panel $(a)$ of Figure 1. This ``1+3" splitting is the \emph{threading} approach to separating time and space of a fundamental observer.

Alternatively, consider \emph{any} congruence of inertial observers. In their bundle of world lines, we identify ``local"  $t$ = constant hypersurfaces, which must represent the \emph{space} experienced by these observers. The vector normal to such a hypersurface must indicate the direction of increasing \emph{time} coordinate. This ``3+1" splitting is the \emph{slicing} approach to separating space and time of an observer family and is schematically depicted in panel $(b)$ of Figure 1. 

To preserve the temporal order of events under Lorentz transformations, we must explicitly exclude superluminal phenomena. To see this, consider the temporal Lorentz transformation from the global inertial frame to another inertial frame moving with velocity ${\mathbf v}$, i.e.,
 \beq \label{1}
t'=\gamma \left(t-\frac{1}{c^2}{\mathbf v}\cdot {\mathbf x}\right)\,.
\eeq
Suppose two events $(t_1,{\mathbf x}_1)$ and $(t_2,{\mathbf x}_2)$ are causally related; hence,  $|\Delta {\mathbf x}|\le c~|\Delta t|$, where
$\Delta {\mathbf x}={\mathbf x}_2-{\mathbf x}_1$ and $\Delta t=t_2-t_1$\,. Thus, $\Delta t'=\gamma \left(\Delta t-{\mathbf v}\cdot \Delta {\mathbf x}/c^2\right)$
implies $\Delta t'/\Delta t>0$, and if $t_2>t_1$, then $t_2'>t_1'$. In this way, causality is preserved if there is no {\it superluminal propagation}.

The treatment of spacetime splitting can be extended to accelerated observers in Minkowski spacetime using the hypothesis of locality, namely, the assumption that an accelerated observer is instantaneously equivalent to an otherwise identical momentarily comoving inertial observer. Further extension to gravitational fields is then possible via Einstein's principle of equivalence. 
The threading and slicing approaches coincide for inertial observers in Minkowski spacetime. Using these extensions to accelerated systems and gravitational fields, we will show that they are compatible with each other when a universal temporal coordinate exists, but are in general \emph{incompatible} in a gravitational field that has no cosmic time.  

In a gravitational field, the threading and slicing approaches lead to restrictions on coordinate systems that may be used to cover the spacetime manifold. The threading coordinate conditions can be invariantly included in the geometric structure of Einstein's theory, so that \emph{all} spacetime coordinate systems must obey these \emph{standard} threading admissibility conditions. The situation is different with the slicing coordinate conditions; in general, they simply impose additional restrictions on coordinate systems so as to exclude causality violation via closed timelike curves. 

It is interesting to study the influence of chronology violation on wave propagation. In fact, Huygens' principle assumes at the outset that coordinate time monotonically increases along the wavefront.  Therefore, wave motion may exhibit some unusual features in chronology violating universes. A complete treatment of this topic is beyond the scope of this paper; however, we explore here some of the consequences of the absence of a cosmic time coordinate for wave motion in a gravitational field. In particular, we study wave propagation parallel to the axis of rotation of stationary G\"odel-type universes. 

Geometrical units ($c=G=1$) will be used hereafter. The plan of this paper is as follows: In Section II, spacetime splitting is discussed in a gravitational field. The resulting conditions on the admissibility of coordinates are examined in Section III. Some aspects of wave motion in certain G\"odel-type universe models are treated in Sections IV - VI. Section VII contains a discussion of our results.

\section{Observers in a Gravitational Field: Threading and Slicing Approaches}

\subsection{Threading Approach}

The {\it fundamental observers} in a gravitational field are those at rest in space. These static observers are naturally connected to a coordinate system. All other observers are pointwise related to the fundamental observers by Lorentz transformations. Imagine a fundamental observer with four-velocity $u$ in an arbitrary gravitational field, as depicted in Figure \ref{fig:3}.  In general, such an observer is accelerated. The spacetime interval along its trajectory---namely, where $dx^i=0$, for $i= 1,2,3$---denotes its proper time, $-ds^2=g_{tt}dt^2$. Therefore, we must have $g_{tt}<0$. Moreover, its orthogonal subspace $\sigma$ represents its local rest space and is such that $u_\mu dx^\mu =0$, where $u^\mu= (-g_{tt})^{-1/2} \delta^{\mu}{}_{0}$. Thus the equation that determines $\sigma$ is $g_{tt}dt+g_{ti}dx^i=0$. In general, one can write
\begin{eqnarray}\label{2}
 g_{\mu \nu}dx^\mu \otimes dx^\nu 
&=&g_{tt}\left(dt+\frac{g_{ti}}{g_{tt}}dx^i \right)\otimes \left(dt+\frac{g_{tj}}{g_{tt}}dx^j \right)+
\left(g_{ij}-\frac{g_{ti}g_{tj}}{g_{tt}}\right)dx^i \otimes dx^j\,,
\end{eqnarray}
so that 
\begin{equation}\label{3}
 g_{\mu \nu}dx^\mu \otimes dx^\nu \big|_\sigma=
\gamma_{ij}dx^i \otimes dx^j\,,
\end{equation}
where 
\beq \label{4}
\gamma_{ij}=g_{ij}-\frac{g_{ti}g_{tj}}{g_{tt}}\,.
\eeq
It follows from the inverse relationship between $(g_{\mu \nu})$ and $(g^{\mu \nu})$ that $(\gamma_{ij})$ and $(g^{ij})$ are inverse of each other; similarly, $(g_{ij})$ and $({\hat \gamma}^{ij})$ are also inverse of each other, where
\beq \label{5}
{\hat \gamma}^{ij}= g^{ij}-\frac{g^{ti}g^{tj}}{g^{tt}}\,.
\eeq
The threading approach thus leads to the following coordinate admissibility conditions
\beq \label{6}
g_{tt}<0\,,\qquad (\gamma_{ij})=\hbox{\rm positive definite matrix}\,.
\eeq
Given a system of coordinates that uniquely identify events in a spacetime region, the standard threading admissibility conditions have to do with the possibility of existence of hypothetical static observers at all events in that region. 

We recall that a real symmetric matrix $\mathcal {A}$ is positive definite if and only if there exists a real invertible matrix $\mathcal {P}$ such that $\mathcal {A} = \mathcal{P}^T\mathcal{P}$. Thus if $\mathcal {A}$ is a positive definite matrix, then so is its inverse; of course, this can also be seen immediately from the positivity of all of their eigenvalues. It follows, for instance,  that for a positive definite matrix $(\mathcal {A}_{ij})$, $\Phi=\mathcal {A}_{ij}dx^i dx^j$ can be expressed in matrix notation as $\Phi=(\mathcal{P}dx)^T(\mathcal{P}dx)$, which is manifestly positive, i.e., $\Phi \ge 0$.

The measurement of spatial distance between a fundamental observer and an arbitrary nearby observer by means of light signals, as well as the synchronization of their clocks, has been discussed in Ref.~\cite{LL}; indeed, it is demonstrated there that the element of spatial distance, $d\ell$, as calculated here, $d\ell^2= \gamma_{ij}dx^i  dx^j$, coincides with the radar distance~\cite{BLM}. 

\subsection{Slicing Approach}

Imagine an \emph{arbitrary} congruence of observers in a gravitational field, as depicted in Figure \ref {fig:4}. This bundle of future directed timelike curves could, for instance, represent the point particles that form a compact object. The ``local" $t=$ constant hypersurface $\Sigma$ is defined by $dt=0$; therefore, 
\beq \label{7}
 g_{\mu\nu}dx^\mu \otimes dx^\nu \big|_{\Sigma}=g_{ij}dx^i\otimes dx^j\,.
\eeq
If $\Sigma$ is to represent space, $(g_{ij})$ must be a positive definite matrix.  Accordingly, its inverse, $({\hat \gamma}^{ij})$, must also be a positive definite matrix. Moreover, the normal to $\Sigma$, ${\mathcal N}_\mu=\delta_\mu{}^0$, must be timelike; hence, $g^{\mu\nu}{\mathcal N}_\mu {\mathcal N}_\nu<0$, or $g^{tt}<0$. Thus the slicing approach leads to the following coordinate admissibility conditions 
\beq \label{8}
g^{tt}<0\,,\qquad ({\hat \gamma}^{ij})=\hbox{\rm positive definite matrix}\,.
\eeq

These conditions ensure that in the coordinate patch under consideration \emph{coordinate time increases monotonically along any timelike curve}. In particular, the application of these conditions to a bundle of fundamental observers implies that in such coordinate systems, $t$ monotonically increases along the world line of a fundamental observer. \emph{Coordinate systems that are admissible according to conditions~\eqref{8} therefore exclude closed timelike curves}.

Let us note that for $g^{tt} \ne 0$, we have in general
\begin{eqnarray} \label{9}
g^{\mu \nu}\partial_\mu \otimes \partial_\nu&=&g^{tt}\left(\partial_t +\frac{g^{ti}}{g^{tt}}\partial_i \right)\otimes \left(\partial_t +\frac{g^{tj}}{g^{tt}}\partial_j \right)
+\left( g^{ij}-\frac{g^{ti}g^{tj}}{g^{tt}}\right) \partial_i \otimes \partial_j\,,
\end{eqnarray}
which should be compared with Eq.~\eqref{2}; in fact, a formal duality between them can be illustrated by means of the lapse and shift functions.  

\begin{figure} 
\typeout{*** EPS figure 1-2}
\begin{center}
\[
\begin{array}{cc}
\includegraphics[scale=0.30]{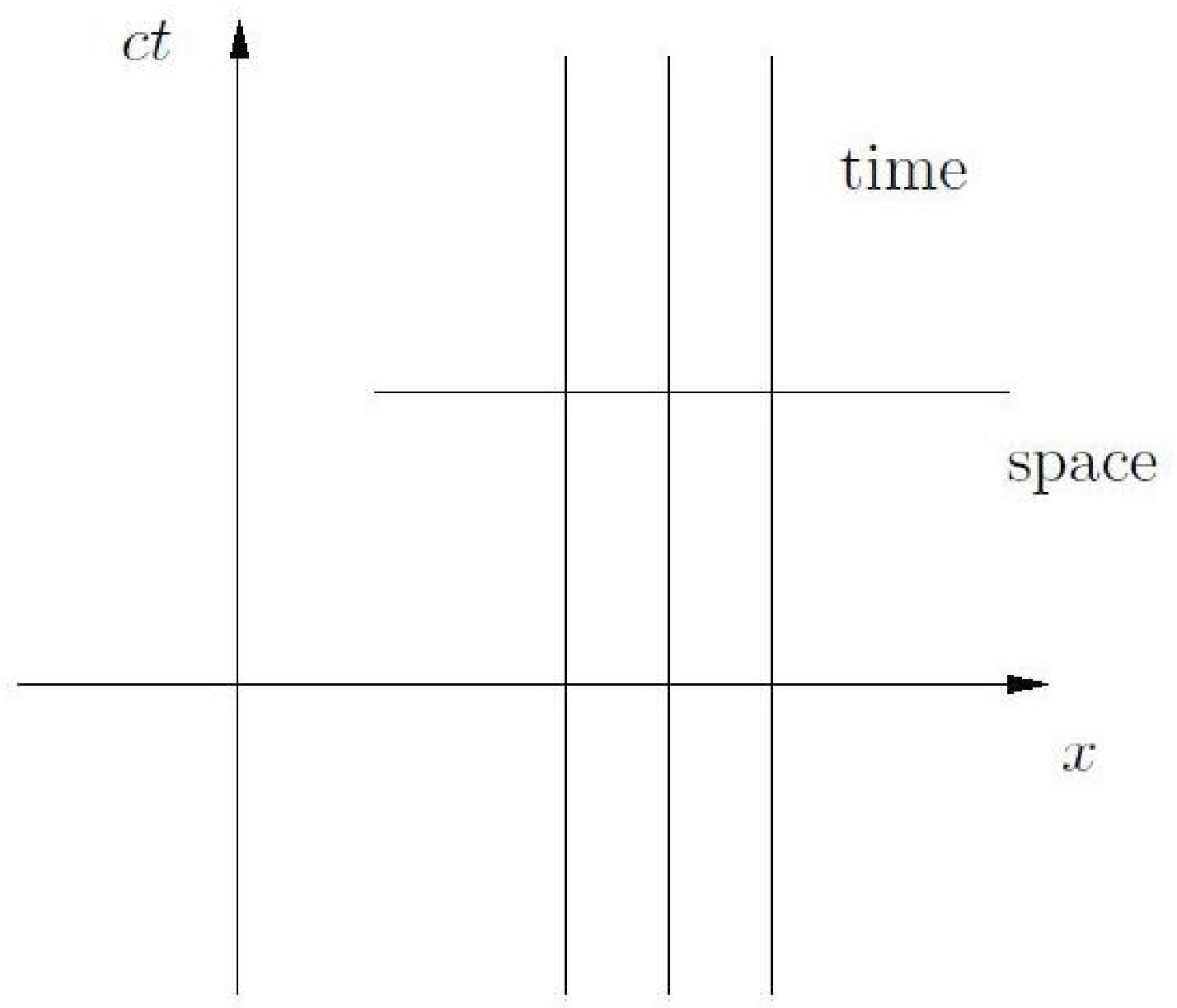} &
\includegraphics[scale=0.30]{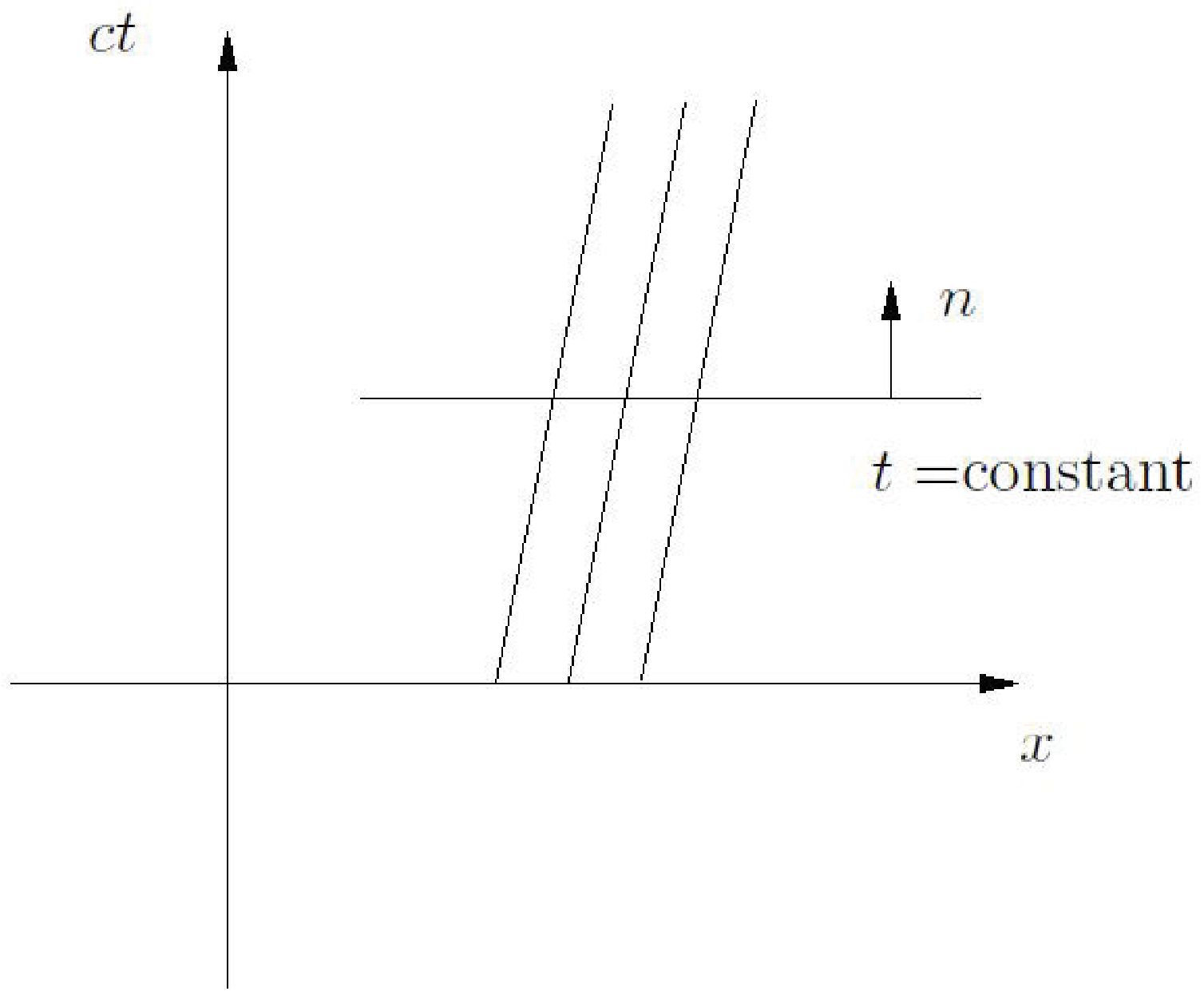} 
\\
(a) &(b) \\
\end{array}
\]
\end{center}
\caption{
(a) Schematic plot representing the {\it threading approach} to spacetime splitting in a global inertial frame in Minkowski spacetime.
(b) Schematic plot representing the {\it slicing approach} for the congruence $x=x_0+vt$, where $x_0$ varies over the congruence and $v$ is a constant speed.}
\label{fig:1-2}
\end{figure} 

\begin{figure}
\typeout{*** EPS figure 3}
\begin{center}
\includegraphics[scale=0.45]{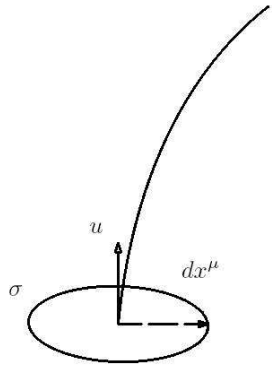}
\end{center}
\caption{Schematic diagram representing the future directed  world line of a fundamental
observer. To every admissible coordinate system, one can associate a set of
fundamental observers, i.e., those at rest in space. The world line of such
an observer is a time line. The infinitesimal orthogonal hypersurface $\sigma$
is not in general a simultaneity hypersurface for the $u$ congruence. The time
lines thread the spacetime; hence, this is the threading approach to splitting spacetime into
time plus space (\lq\lq 1+3").}
\label{fig:3}
\end{figure}

\begin{figure}
\typeout{*** EPS figure 4}
\begin{center}
\includegraphics[scale=0.45]{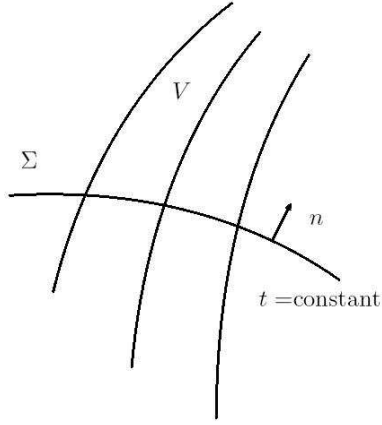}
\end{center}
\caption{Consider any congruence of future directed timelike world lines $V$ in a standard
admissible coordinate system. The infinitesimal constant-time hypersurfaces
$\Sigma$ should all be spacelike. Thus their normal vector field should be
timelike. The hypersurfaces slice up this spacetime region; therefore, this is the slicing approach
to the splitting of spacetime into space plus time (\lq\lq 3+1"). Note that $n_\mu V^\mu >0$ and the temporal coordinate monotonically increases along such a congruence. } 
\label{fig:4}
\end{figure}

\subsection{Duality}

In the threading approach, the lapse and shift functions are respectively $M$ and $M_i$ given by   
\beq
M=\sqrt{-g_{tt}}\,,\qquad M_i=-\frac{g_{ti}}{g_{tt}}\,.
\eeq
Then, we have for  metric~\eqref{2} the equivalent representation 
\beq
\label{thd2}
-ds^2=-M^2\left(dt-M_idx^i \right)\left(dt-M_jdx^j \right)+\gamma_{ij} dx^i dx^j\,,\qquad \gamma_{ij}=g_{ij}+M^2M_iM_j\,.
\eeq
The form of  metric (\ref{thd2}) is adapted to static observers with $4$-velocity
\beq
u=\frac{1}{M}\partial_t\,,\qquad u^\flat =-M(dt-M_idx^i)\,,
\eeq
where the ``flat" symbol ($\flat$) denotes the fully covariant form of a tensor. In fact,
\beq
\label{thd3}
-ds^2=-(u_\alpha dx^\alpha)^2+\gamma_{ij} dx^i dx^j\,.
\eeq

In a similar way,  one may introduce the lapse - shift notation in the slicing approach, namely, 
\beq
N=\frac{1}{\sqrt{-g^{tt}}}\,,\qquad N^i=-\frac{g^{ti}}{g^{tt}}\,.
\eeq
Normalizing ${\mathcal N}_\mu$, so that it becomes a  vector of unit length $n_\mu$, we have 
\beq
n=-\frac{1}{N}(\partial_t -N^i \partial_i)\,, \quad  n^\flat =Ndt\,.
\eeq
Thus in the slicing approach, Eq.~\eqref{9} can be expressed in terms of $n^\alpha$ and ${\hat\gamma}^{ij}$ in a manner that is completely analogous to the threading approach, namely,
\begin{eqnarray}
-(\partial_s)^2&=& -(n^{\alpha} \partial_{\alpha})^2 + {\hat \gamma}^{ij} ( \partial_i) ( \partial_j) .
\end{eqnarray}
However, a closer examination reveals that this formal duality does \emph{not} extend to the physical conditions contained in Eqs.~\eqref{6} and~\eqref{8}.

At any event $p$ on the spacetime manifold, we have a tangent space $T_p$ of vectors at $p$ as well as the dual space $T^*_p$ of one-forms at $p$ with corresponding expressions of the metric tensor given respectively by Eqs.~\eqref{2} and~\eqref{9}. At $p$, the symmetric matrix $(g_{\mu \nu})$ can be diagonalized, so that $(g_{\mu \nu})=Q^{-1}DQ$, where $Q$, $QQ^T=1$, is an orthogonal matrix and $D={\rm diag}\, (d_0,d_1,d_2,d_3)$, where $d_0<0$ and $d_i>0$, for $i=1,2,3$. Thus by means of a local transformation and a certain scaling, $(g_{\mu \nu})$ at each point $p$ can be reduced to $(\eta_{\mu \nu})$. Furthermore, we have that  at $p$, $(g^{\mu \nu})=Q^{-1}D^{-1}Q$, where the diagonal elements of $D^{-1}$ respect the metric signature, as before. In a similar way, $(g^{\mu \nu})$ can be reduced to $(\eta^{\mu \nu})$ at $p$. This pointwise duality thus stays at the threading level and does \emph{not} extend to the slicing approach. That is, the conditions given in Eq.~\eqref{8} are in general different from those given in Eq.~\eqref{6}, as can be seen via explicit examples discussed in Section IV.

\section {Admissibility Conditions}

The coordinate conditions~\eqref{6} that result from the threading approach can be naturally incorporated into the underlying geometric framework of general relativity; however, the slicing conditions~\eqref{8} can be neither invariantly formulated nor are they sufficiently local to become part of the structure of Lorentzian geometry. Indeed, all coordinate patches in which the metric takes the form $g_{\mu \nu}(x)= \eta_{\mu \nu}+ h_{\mu \nu}(x)$, where $h_{\mu \nu}(x)$ can be treated as a small perturbation---such as in the Riemann normal coordinates about any event or the Fermi normal coordinates about the world line of an observer---satisfy both conditions~\eqref{6} and~\eqref{8}. While coordinate systems that are compatible with conditions~\eqref{8} protect chronology, this does not mean that chronology violation does not occur in the gravitational field under consideration. In fact, general relativity per se does not require the existence of a cosmic time, as conditions~\eqref{8} cannot be woven into the fabric of Lorentzian geometry \cite{Beem}. From this viewpoint, general relativity must be abandoned if one insists on integrating the slicing coordinate conditions into the foundations of the theory of gravitation.            

The threading and slicing approaches and their respective restrictions on coordinate charts have been discussed in one form or another by a number of authors. We should mention in particular the work of Zelmanov~\cite{ZEL, ZEL2}, Cattaneo~\cite{CAT}, M{\o}ller~\cite{MOL} and Lichnerowicz~\cite{LICH}, among others. A historical review is beyond the scope of this paper. Following Synge~\cite{SYN}, we refer to the combined threading and slicing conditions as the Lichnerowicz admissibility requirements (see the theorem on page 9 of Ref.~\cite{LICH}). The book of Landau and Lifshitz~\cite{LL} contains a particularly clear discussion of the standard admissibility requirements. 

It is a theorem of linear algebra~\cite{HK} that a real symmetric $n\times n$ matrix $A$ is {\it positive definite} if and only if the {\it principal minors} of $A$ are all positive. These are $n$ scalars defined by
\beq
{\rm det} \left[ 
\begin{array}{ccc} 
A_{11}& \ldots & A_{1k}\cr
. & & . \cr
. & & . \cr
A_{k1}& \ldots & A_{kk}\cr
\end{array}
\right]\,,\qquad k=1,\ldots, n\,.
\eeq
Now consider the standard threading admissibility conditions.
These involve the matrix $(g_{\mu\nu})$: $g_{tt}<0$ and $\gamma_{ij}=g_{ij}-g_{ti} g_{tj} / g_{tt}$ must form a positive definite matrix. Therefore,
\beq
(\gamma_{ij})=\, \left[ 
\begin{array}{ccc} 
\gamma_{11}& \gamma_{12}& \gamma_{13}\cr
\gamma_{12}& \gamma_{22}& \gamma_{23}\cr
\gamma_{13}& \gamma_{32}& \gamma_{33}\cr
\end{array}
\right]\,
\eeq
must be such that
\beq
\gamma_{11}>0\,,\qquad \gamma_{11}\gamma_{22}-\gamma_{12}^2>0\,,\qquad {\rm det}\, (\gamma_{ij})>0\,.
\eeq
Note that $g_{tt}<0$ and $\gamma_{11}=g_{11}-g_{t1}^2/g_{tt}>0$ together imply that
\beq
g_{tt}~\gamma_{11}={\rm det} \left[ 
\begin{array}{cc} 
g_{tt}& g_{t1}\cr
g_{1t}&  g_{11}\cr
\end{array}
\right]<0\,.
\eeq
Next, $\gamma_{11}\gamma_{22}-\gamma_{12}^2>0$, when written out in detail, can be expressed as
\beq
{\rm det} \left[ 
\begin{array}{cc} 
\gamma_{11}& \gamma_{12}\cr
\gamma_{21}&  \gamma_{22}\cr
\end{array}
\right]=\frac{1}{g_{tt}}~
{\rm det} \left[ 
\begin{array}{ccc} 
g_{tt}& g_{t1} & g_{t2}\cr
g_{1t}&  g_{11}& g_{12}\cr
g_{2t}&  g_{21}& g_{22}\cr
\end{array}
\right]\,.
\eeq
Similarly, one finds that (see Appendix A)
\beq
\label{A6}
{\rm det}(\gamma_{ij})=\frac{1}{g_{tt}}~{\rm det} (g_{\mu\nu})\,.
\eeq

Putting all these results together, we find that \emph{the standard threading admissibility conditions are equivalent to the statement that the principal minors of
$(g_{\mu\nu})$ must all be negative for our $(-,+,+,+)$ signature. That is $(g_{\mu\nu})$ must be a negative definite matrix. Similarly, the slicing admissibility conditions imply that $(g^{\mu\nu})$ must be a negative definite matrix.
The Lichnerowicz admissibility conditions then require that both $(g_{\mu\nu})$ and $(g^{\mu\nu})$ be negative definite matrices, i.e., all their principal minors must be negative.}

In many spacetimes of interest in general relativity, Lichnerowicz  conditions reduce to the standard admissibility conditions; indeed, most solutions of Einstein's equations are of this type~\cite{ES}. This is the case, for instance, for the Kerr metric in the standard Boyer-Lindquist coordinates \emph{outside the static limit}; however, CTCs occur in the region interior to the inner event horizon. Moreover, geodesic coordinate systems established along the world line of an accelerated observer in Minkowski spacetime are such that the threading and slicing conditions coincide~\cite{Bahram, bahram_ferrarese}; somewhat similar results are expected for \emph{radar coordinates}~\cite{BLM}. The important point here is that Lichnerowicz  conditions exclude CTCs, whereas the standard coordinate conditions do not in general exclude them. Much has been written about CTCs and time travel in general relativity---see, for example,~\cite{bonnor1969, bonnor2001, bonnor2003, cooperstock}, the recent review~\cite{Lobo} and the references cited therein.

\section{Causality and Wave Propagation}

Consider the propagation of test radiation in a coordinate patch containing a CTC. The CTC might leave an imprint on waves propagating in such a spacetime domain and such a trace could then be observationally detectable. In this connection, it appears useful to study wave phenomena in spacetimes with CTCs. Consider, for example, a massless field propagating in a spacetime region with a \emph{closed null geodesic}. In the WKB approximation, we can imagine a wave packet moving essentially along the closed null geodesic, since in the WKB limit, massless wave propagation reduces to null geodesic motion.  An examination of this situation reveals that the wave would appear to get immediately reflected: if the wave is initially moving forward in time, its spatial direction of propagation gets reversed when it starts to move backward in time. A similar result is expected for massive fields. We recall that a wave with propagation vector $k^\mu=(\omega, \mathbf{k})$ in an inertial frame of reference is, up to a certain amplitude function, of the form $\exp(-i\omega t+i\mathbf{k}\cdot \mathbf{x})$, so that as $t$ increases, $\mathbf{x}$ increases along the wave vector $\mathbf{k}$ as the wave travels forward; however, if $t$ decreases, the wave travels in the $-\mathbf{k}$ direction. When the direction of the arrow of time reverses, the wave appears to suffer a reflection. Thus we expect some unusual features for wave phenomena in gravitational fields that do not have a global time coordinate.

Consider, for instance, the propagation of waves in the exterior Kerr spacetime~\cite{matzner}; in fact, there is absorption of radiation through the horizon in this case, but the scattering process contains little information about the CTCs that exist in the \emph{interior} Kerr spacetime.  Thus we must concentrate on  wave propagation in regions where the CTCs are accessible. An ideal case would be the G\"odel universe~\cite{God, Ellis, HE, NDT}, which is free of horizons and singularities. Do  waves propagating in the G\"odel universe exhibit such an imprint~\cite{bm, KoOb, S1, S2}? In fact, the elucidation of this issue originally motivated the preliminary analysis presented in this paper and, as will become clear below, its complete treatment remains a task for the future. 

The metric of G\"odel's universe in the original quasi-Cartesian coordinates can be written as
\beq
\label{godel_cart}
-ds^2=-dt^2-2\sqrt{2}Udtdy+dx^2-U^2 dy^2+dz^2\,,\qquad U=e^{\sqrt{2}\Omega x}\,,
\eeq 
where $\Omega>0$ is the frequency of universal rotation and $\partial_t$, $\partial_y$ and $\partial_z$ are three of the five Killing vectors of this spacetime. The determinant of the metric turns out to be $g=-U^2$ and the inverse metric is thus given by
\beq
\label{invGodel}
-(\partial_s)^2=(\partial_t)^2-2\sqrt{2}U^{-1}(\partial_t)(\partial_y)+(\partial_x)^2+U^{-2}(\partial_y)^2+(\partial_z)^2\,.
\eeq
The fundamental observers follow geodesics of metric~\eqref{godel_cart}; in fact, their four-velocity $u$ is given by $u=\partial_t$, so that we have
\begin{eqnarray}
\qquad
G_{\alpha\beta}&=&\Omega^2\left( g_{\alpha\beta}+2u_\alpha u_\beta \right)\,.
\end{eqnarray}
The source of G\"odel's solution of the Einstein field equations is dust of  constant energy density $\rho$, with fluid  unit four-velocity  aligned with the time coordinate lines, and cosmological constant $\Lambda=-\Omega^2=-4\pi\rho$. Here  $\Omega>0$ describes an intrinsic counterclockwise rotation of the universe around the $z$ axis. The threading conditions are satisfied by metric~\eqref{godel_cart}, while it is clear from the inverse metric~\eqref{invGodel} that $g^{tt}=1$. Indeed, the slicing coordinate conditions do not hold  and the original G\"odel coordinates are therefore not admissible in the sense of Lichnerowicz.

In~\cite{chimas}, exact Fermi normal coordinates were first explicitly introduced along the world line of a fiducial fundamental observer in G\"odel spacetime. Imposing Lichnerowicz conditions on these Fermi coordinates limits their domain of applicability to the interior of a cylindrical region of radius $\sqrt{2} \ln (1+\sqrt{2})/{\Omega} \simeq {1.25}/{\Omega}$ about the axis of rotation~\cite{chimas}. This boundary cylinder consists of closed null curves beyond which CTCs exist. 

Propagation of test electromagnetic radiation in the G\"odel universe was first investigated in~\cite{bm} with the express purpose of demonstrating the coupling of photon spin with the essentially uniform gravitomagnetic field of this rotating universe. The formal simplicity of the G\"odel metric made it possible to find the exact solution of the perturbation equations and thus establish the gravitational coupling of the helicity of the photon to the rotation of the G\"odel universe. To solve Maxwell's equations, it was deemed convenient to use the Skrotskii method, which is based on the analogy between a gravitational field and an optical medium~\cite{Sk}, and replace the G\"odel field with an effective gyrotopic medium. As is well known, in the treatment of electrodynamics in arbitrary spacetime coordinates, it is always possible to introduce instead a certain hypothetical medium with definite constitutive properties that occupies Minkowski spacetime in quasi-Cartesian coordinates~\cite{Fel}; in fact, this method is conceptually related to Weyl's formulation of Fermat's principle in general relativity and to Gordon's optical metric~\cite{SYN, Per, HYu}. The physical description of photon helicity states is straightforward in scattering situations involving asymptotically flat spacetimes, but requires special care in universe models that are not asymptotically flat. For instance, in the G\"odel universe this connection is established via the limit as $\Omega \to 0$ and G\"odel spacetime reduces to Minkowski spacetime. Furthermore, we expect that the spectrum of electromagnetic wave perturbations of the G\"odel universe would contain certain discreteness properties in close analogy with the Landau spectrum of charged particles in a uniform magnetic field. Out of the complete spectrum of finite perturbations, there is a part that satisfies these various physical requirements and only this part of the full spectrum was given in ~\cite{bm} without providing details of the selection process. The preferred part of the spectrum involves  wave vectors whose magnitudes are sufficiently large compared to $\Omega$, corresponding to a pulse of radiation produced by a localized source. From this solution one finds, in addition to the spin-rotation-gravity coupling, that \emph{waves cannot propagate parallel to the direction of rotation of this universe} and, moreover,  that waves can propagate only in one direction along the $y$ axis. We wish to determine how these features as well as the spectrum are connected with the violation of causality in the G\"odel universe.  

Electromagnetic fields in the G\"odel universe have been discussed via the Debye potential formalism in~\cite{CVD}. Moreover, the scalar and neutrino perturbations of the G\"odel universe have been studied by a number of authors---see~\cite{WAH, SNGT, DAL, PM, PCM} and the references therein.

To see the imprint of acausality on these wave results, it is necessary to compare them with the results of wave propagation in causal G\"odel-type rotating universes that possess cosmic time. Indeed, G\"odel-type solutions of Einstein's equations exist that contain both causal and acausal universe models---see~\cite{RT, OK, KO, Yuri} and the references cited therein. The metric of such G\"odel-type solutions is of the form 
\beq
\label{g-t1}
-ds^2=-dt^2-2\eta R(t)\mathcal{U} dtdy+R^2(t)[dx^2-(\eta^2-1) \mathcal{U}^2 dy^2+dz^2]\,,
\eeq 
where $R(t)>0$ is the scale factor,
\beq
\label{g-t2}
\mathcal{U}=e^{\lambda x}\,, \qquad \eta=\frac{2\Omega}{\lambda}\,.
\eeq 
Here $\lambda>0$ is a constant parameter and $\Omega > 0$ is a vorticity parameter. The G\"odel solution is recovered for $R=1$, $\lambda=\sqrt{2} \Omega$ and $\eta=\sqrt{2}$.  In general, Eq.~\eqref{g-t1} represents a spacetime of Petrov type D with three Killing vector fields given by $\partial_x-\lambda y \partial_y$, $\partial_y$ and $\partial_z$, so that it is a spatially homogeneous universe of type III in the Bianchi classification. Let $\theta^\mu =  \delta^{\mu}{}_{0}$ be the four-velocity vector of the family of fundamental observers in the G\"odel-type universe, then this congruence has expansion and rotation, but no shear. In particular, the \emph{rotation tensor} can be expressed as
\beq
f_{\alpha\beta}=\theta_{[\alpha;\beta]}+a_{[\alpha}\theta_{\beta ]}\,,
\eeq
where $a_\mu$ is the acceleration of this congruence given by $a^\flat =-\eta \dot R {\mathcal U} dy$ with $\dot R = dR/dt$. It follows that the only nonzero components of $f_{\alpha\beta}$ are given by $f_{xy}=-f_{yx}=\Omega R{\mathcal U}$, so that the vorticity of the congruence of the fundamental observers can be expressed as
\beq
{\tilde \Omega}=\Big(\frac12 f_{\alpha\beta}f^{\alpha\beta}\Big)^{1/2}=\frac{\Omega}{R(t)}\,.
\eeq
Only positive square roots are considered throughout this paper. Moreover, the expansion is given by $3\dot R/R$. In the special case of stationary G\"odel-type universes with $R=1$, the fundamental observers follow geodesics and the vorticity of their congruence is simply $\Omega$. Various sources have been considered for universes of G\"odel type~\cite{RT, OK, Yuri}; moreover, the corresponding two-fluid cosmological models have been discussed in~\cite{Dunn}.

For metric~\eqref{g-t1}, we find $g=-R^6\mathcal{U}^2$ and hence the inverse metric is given by
\beq
\label{g-t3}
-(\partial_s)^2=(\eta^2-1)(\partial_t)^2-2\eta R^{-1}\mathcal{U}^{-1}(\partial_t)(\partial_y)+R^{-2}[(\partial_x)^2+\mathcal{U}^{-2}(\partial_y)^2+(\partial_z)^2]\,.
\eeq
The threading coordinate conditions are satisfied here, while the slicing coordinate conditions are only satisfied for $\eta<1$. Thus the solutions are \emph{causal} for $\eta<1$ and \emph{acausal} for $\eta>1$, while $\eta=1$ constitutes a rather special limiting case. 

A detailed treatment of the propagation of electromagnetic waves in this G\"odel-type universe is due to Korotky and Obukhov~\cite{KoOb} and Saibatalov~\cite{S1}, who provided a complete description of the spectrum and outlined the main differences between the spectra in the causal and acausal cases. We emphasize that the more extended treatments of the G\"odel case in~\cite{KoOb} and~\cite{S1} go beyond the results of~\cite{bm}, where only the part of the spectrum that was deemed most physically reasonable was presented. In particular, the discrepancy between the results of~\cite{bm} and~\cite{S1} is \emph{not} due to any inherent limitation of the Skrotskii method.

It is interesting to note that scalar and neutrino perturbations of G\"odel-type universes have been the subject of recent investigations~\cite{S2, Mar, KZ}. 

In the following section, we investigate the propagation of radiation parallel to the direction of rotation of a \emph{stationary} G\"odel-type universe with $R=1$. This two-parameter subclass has been thoroughly investigated in~\cite{TRA, RA} and the references cited therein.

\section{Waves in Stationary G\"odel-Type  Universes}

Consider the propagation of waves in metric~\eqref{g-t1} with $R=1$. It turns out that the main  result of this analysis is essentially independent of any particular wave equation; therefore, we consider the scalar wave equation for the sake of simplicity. We seek a solution of
\beq
\label{g-t4}
\nabla_\alpha \partial^\alpha \Psi-{\hat m}_0^2\Psi=0\, 
\eeq
in the form
\beq
\label{g-t5}
\Psi=e^{i(-\omega t +k_2 y+k_3 z)}\psi(x)\,,
\eeq
where ${\hat m}_0=m_0 / \hbar$, $m_0$ is the mass of the scalar particle and $\omega$, $k_2$ and $k_3$ are constant propagation parameters. The function $\psi(x)$ satisfies
\beq
\label{g-t6}
\mathcal{U}\frac{d}{dx}\Big(\mathcal{U} \frac{d\psi}{dx}\Big) - \lbrace k_2^2+2\eta \omega k_2~ \mathcal{U} + [(\eta^2-1)\omega^2+k_3^2+{\hat m}_0^2]~\mathcal{U}^2\rbrace\psi=0\,. 
\eeq
Now introducing $\mathcal{U}^{-1}=\chi$ as a new variable and imposing the requirement that the wave travel in the $z$ direction ($k_2=0$), we find
\beq
\label{g-t7}
\lambda^2 \chi^2~ \frac{d^2\psi}{d\chi^2}- [(\eta^2-1)\omega^2+k_3^2+{\hat m}_0^2] \psi=0\,. 
\eeq
Alternatively, Eq.~\eqref{g-t6} with $k_2=0$ can be written in terms of variable $x$ as a second order ordinary differential equation with constant coefficients and its solutions are of the form $\exp(-\lambda \sigma x)$. For $x: -\infty \to \infty$, $\chi: \infty \to 0$ and the solutions for $\psi$ can be written as $\chi^{\sigma}$, where
\beq
\label{g-t8}
\lambda^2 \sigma(\sigma-1)= (\eta^2-1)\omega^2+k_3^2+{\hat m}_0^2\,. 
\eeq
The only finite solution for $\psi$ in the range $\chi: 0 \to \infty$ is a \emph{constant}, provided 
\beq
\label{g-t9}
(\eta^2-1)\omega^2+k_3^2+{\hat m}_0^2=0\,. 
\eeq
Therefore, if $\eta<1$, so that the stationary G\"odel-type universe is \emph{causal}, we have
\beq
\label{g-t10}
\Psi=\psi_0 e^{i(-\omega t + k_3 z)}\,, \qquad \omega^2=\frac{k_3^2+{\hat m}_0^2}{1-\eta^2}\,,
\eeq
where $\psi_0$ is a constant; otherwise, for $\eta \ge 1$ wave propagation parallel to the rotation axis is impossible. That is, waves can only propagate parallel to the rotation axis of \emph{causal} stationary G\"odel-type universes. Let us recall here that waves can freely propagate parallel to the rotation axis of a compact gravitating source as well as a Kerr black hole~\cite{matzner, bm, Sk}.

What happens to a packet of scalar radiation produced in the neighborhood of $z=z_0$ in the acausal ($\eta \ge 1$) G\"odel-type universe? It follows from Eq.~\eqref{g-t9} that the radiation is then confined in space around $z_0$. That is, with $k_3=\pm i\kappa$, where we use the upper sign for $z>z_0$ and the lower sign for $z<z_0$, $\kappa > 0$ is given by
\beq
\label{g-t9'}
\kappa=\Big[(\eta^2-1)\omega^2+{\hat m}_0^2\Big]^{1/2}\,. 
\eeq
Then the solution for the scalar field can be expressed as
\beq
\label{g-t9''}
\Psi=\phi_0 e^{- i\omega t - \kappa (z-z_0)}\,, \qquad z \ge z_0 \,,
\eeq
and
\beq
\label{g-t9'''}
\Psi=\phi_0 e^{- i\omega t + \kappa (z-z_0)}\,, \qquad z \le z_0 \,,
\eeq
where $\phi_0$ is a constant. 

Consider now the propagation of radiation in the limit of high frequencies and wave numbers. For electromagnetic radiation, the geometric optics (or eikonal) approximation scheme results in the massless Hamilton-Jacobi equation and the corresponding propagation along null geodesic rays~\cite{BM}. This is indeed the case for any reasonable massless wave equation. In the massive case, the analogous treatment is the WKB approximation, where for the scalar field we set
\beq
\label{g-t10'}
\Psi=e^{iS/\hbar} (\Psi_0 + \hbar \Psi_1 + ...).
\eeq
Substitution of this ansatz into Eq.~\eqref{g-t4} results in an expansion in increasing powers of $\hbar$ starting with $\hbar^{-2}$; for $\hbar \to 0$, the terms in the expansion proportional to $\hbar^{-2}$, $\hbar^{-1}$ and $\hbar^{0}$ must vanish. The primary result, which follows from setting the coefficient of $\hbar^{-2}$ equal to zero, is indeed independent of the nature of the wave 
and is the Hamilton-Jacobi equation
\beq
\label{g-t11}
g^{\mu \nu}\nabla_\mu S\nabla_\nu S +m_0^2=0\,,
\eeq
where the scalar field $S$ is related to action. The other equations then describe, for instance, how the wave amplitude $\Psi_0$ propagates along the geodesic that is the solution of the Hamilton-Jacobi equation, and so on~\cite{BM}.

In the case under consideration, Eq.~\eqref{g-t3} implies that 
\beq
\label{g-t12}
(\eta^2-1)\Big(\frac{\partial S}{\partial t}\Big)^2-2 \eta \mathcal{U}^{-1}\Big(\frac{\partial S}{\partial t}\Big)\Big(\frac{\partial S}{\partial y}\Big)+\Big(\frac{\partial S}{\partial x}\Big)^2+\mathcal{U}^{-2}\Big(\frac{\partial S}{\partial y}\Big)^2+\Big(\frac{\partial S}{\partial z}\Big)^2+m_0^2          =0\,. 
\eeq
For propagation along the $z$ direction, it is natural to assume that  
\beq
\label{g-t13}
S=-Et+P_3z\,, 
\eeq
where $E$ and $P_3$ are real constants representing the energy and momentum of the particle. Substitution of Eq.~\eqref{g-t13} into Eq.~\eqref{g-t12} yields
\beq
\label{g-t14}
(\eta^2 -1) E^2 + P_3^2 +m_0^2=0\,. 
\eeq
This equation has no real solution in the acausal case ($\eta \ge 1$), but in the causal case ($\eta<1$), the solution corresponds to Eq.~\eqref{g-t10}, as expected. That is, the wavefront is the hypersurface of constant phase, which is proportional to $S$ in the eikonal approximation; therefore, hypersurfaces of constant $S$ must be spacelike (for massive fields) or null (for massless fields) in accordance with the Hamilton-Jacobi equation.

In general, the solution of the Hamilton-Jacobi equation corresponds to a geodesic. To see this, let $m_0 w_{\mu}=\nabla_{\mu}S$ and note that $w_{\mu;\nu}=w_{\nu;\mu}$; that is, $w_{\mu}$ is a curl-free vector field. Taking covariant derivative of Eq.~\eqref{g-t11}, we find that $g^{\mu \nu}w_{\mu;\alpha}w_{\nu}=0$. This can be written, using the curl-free property, as $g^{\mu \nu}w_{\alpha;\mu}w_{\nu}=0$, which is the geodesic equation for $w^{\mu}=dx^{\mu}/ds$. Thus in the case under consideration, $m_0 w_{\mu}=(-E, 0, 0, P_3)$ and one can easily find the curl-free geodesic congruence in the causal case ($\eta<1$), namely, 
\beq
\label{g-t15}
t=t_0+\frac{E}{m_0}(1-\eta^2)(\tau-\tau_0)\,,
\eeq
where $\tau$ is the proper time along the world line and
\beq
\label{g-t15'}
 x=x_0, \quad y=y_0+\frac{\eta E}{m_0~ \mathcal{U}(x_0)}(\tau-\tau_0), \quad  z=z_0+\frac{P_3}{m_0}(\tau-\tau_0)\,. 
\eeq
Here $t_0, x_0, y_0, z_0$ and $\tau_0$ are constants of integration and $E$ and $P_3$ are connected via Eq.~\eqref{g-t14}. There is no corresponding geodesic for $\eta \ge 1$, as the Hamilton-Jacobi equation results in an imaginary energy in this case. 

It is puzzling that the geodesic corresponding to the eikonal limit of waves propagating in the $z$ direction for $\eta<1$ involves motion in the $y$ direction as well. Let us note that in general, metric~\eqref{g-t1} allows geodesics moving parallel to the axis of rotation \emph{regardless of the value of $\eta$}; indeed, 
\beq
\label{g-t16}
\zeta^\mu=\gamma(1, 0, 0, \beta)\,, 
\eeq
where $\beta$ and $\gamma=(1-\beta^2)^{-1/2}$ are constants, is the unit four-velocity of such geodesic observers. However, $m_0 \zeta_\mu$ is \emph{not} the gradient of a scalar, since $\zeta_\mu$ is not a curl-free vector field, as can be directly verified. A bundle of neighboring geodesics has in general nonzero vorticity; on the other hand, if the vorticity vanishes at one point, then the bundle will be completely free of vorticity---see page 84 of~\cite{HE}. 

The remaining difficulty has to do with the fact that the Hamilton-Jacobi equation does not result in a \emph{real} particle energy for $\eta \ge 1$. \emph{This circumstance is a consequence of the violation of slicing coordinate conditions in metric~\eqref{g-t1}.} To see this in general, note that the geodesic equation is derived from the extremum of the action, $\delta S=0$, where
\beq
\label{g-t17}
S=-m_0 \int ds\, 
\eeq
is a functional of the path that connects the two \emph{fixed} events that are endpoints of the integral in Eq.~\eqref{g-t17}. Let us define the Lagrangian $L(t, x^i, v^j)=-m_0 ds/dt$, where $v^i=dx^i/dt$. Using $p_i=\partial L / \partial v^i$, the corresponding Hamiltonian $H(t, x^i, p_j)=p_iv^i-L$ works out to be $-p_0$, where $p_0$, as a function of $t, x^i$ and $p_j$, must be obtained from $g^{\mu \nu}p_\mu p_\nu+m_0^2=0$. This equation has the solution
\beq
\label{g-t18}
-p_0=H(t, x^i, p_j)=\frac{g^{ti}}{g^{tt}}p_i  \pm\Big(\frac{m_0^2+{\hat \gamma}^{ij}p_ip_j}{-g^{tt}}\Big)^{1/2}\,. 
\eeq
\emph{Thus the slicing coordinate conditions would ensure that the Hamiltonian in Eq.~\eqref{g-t18} is real}~\cite{ChMa}. Then, integrating along \emph{geodesics} in Eq.~\eqref{g-t17}, but now with only the initial event kept fixed and the final event considered variable, we have
\beq
\label{g-t19}
S(t, x^i)= \int^{(t, x^i)} (p_j v^j - H)dt\,, 
\eeq
so that $dS=p_idx^i+p_0dt$, or $\nabla_\mu S=p_\mu$, which then leads directly to the Hamilton-Jacobi equation. However, if Eq.~\eqref{g-t18} cannot produce a real solution, then a real Hamiltonian does not exist and the connection of the geodesic equation with the Hamilton-Jacobi equation and hence wave propagation is severed.  We should mention in passing that this derivation of the Hamilton-Jacobi equation implies that a fountain of neighboring geodesics emanating from a single event has vanishing vorticity. 

It is interesting to illustrate these results for the propagation of Dirac particles parallel to the rotation axis of stationary G\"odel-type universes.

\section{Dirac Equation in Stationary G\"odel-Type  Universes}

Let us now consider the massive Dirac equation in a stationary G\"odel-type universe described by  metric~\eqref{g-t1} with $R=1$. Here we follow the standard notation for the Newman-Penrose formalism as employed in the monograph of Chandrasekhar~\cite{chandra}. Therefore, due to the nature of the subject matter, the notations and conventions used in this section are generally independent of the rest of the paper, except when otherwise indicated; in particular, we switch the sign of the metric signature.

An orthonormal frame naturally adapted to the spacetime coordinates is given by
\begin{eqnarray}\label{Dirac1}
\omega^{\hat t}=dt +  \eta {\mathcal U} dy\,,\quad
\omega^{\hat x}=  dx\,,\quad
\omega^{\hat y}=   {\mathcal U} dy\,,\quad
\omega^{\hat z}=  dz,
\end{eqnarray}
which can be used to form a null tetrad frame
\beq
l=\frac{1}{\sqrt{2}}(\omega^{\hat t}+\omega^{\hat z})\,,\quad
n=\frac{1}{\sqrt{2}}(\omega^{\hat t}-\omega^{\hat z})\,,\quad
m=\frac{1}{\sqrt{2}}(\omega^{\hat x}+i\omega^{\hat y})\,.
\eeq
The associated non-vanishing spin coefficients are
\beq
\alpha =-\beta=-\frac{\sqrt{2}}{4}\lambda\,,\quad  
\mu =\rho=2\epsilon=2\gamma=-\frac{i\sqrt{2}}{2}\Omega\,, 
\eeq
while the only nonvanishing Weyl scalar is
\beq
\psi_2 = -\frac{\lambda^2}{6} (\eta^2-1)\,,
\eeq
as metric~\eqref{g-t1} is of Petrov type D.
Following \cite{chandra}, the massive Dirac  equation in the Newman-Penrose formalism is summarized by the following set of equations
\begin{eqnarray}
(D+\epsilon -\rho)F_1 +(\delta^*+\pi-\alpha)F_2 &=&i m_* \, G_1, \nonumber\\
(\Delta +\mu -\gamma)F_2 +(\delta+\beta-\tau)F_1 &=& i m_* \, G_2, \nonumber\\
(D+\epsilon^* -\rho^*)G_2 -(\delta+\pi^*-\alpha^*)G_1 &=& i m_* \, F_2, \nonumber\\
(\Delta +\mu^* -\gamma^*)G_1 -(\delta^*+\beta^*-\tau^*)G_2 &=& i m_* \, F_1\,,
\end{eqnarray}
where $F_1,F_2,G_1$ and $G_2$ are the components of the Dirac spinor, $m_*={\hat m}_0/\sqrt{2}$ is proportional to the mass of the Dirac particle (see \cite{chandra}, p. 543) and
\beq
D=l^\mu \nabla_\mu \,,\quad \Delta=n^\mu \nabla_\mu \,,\qquad \delta=m^\mu \nabla_\mu 
\eeq
are directional derivatives along the null tetrad frame.
Thus,
\begin{eqnarray}
(D+\epsilon -\rho)F_1 +(\delta^*-\alpha)F_2 &=&i m_* \, G_1, \nonumber\\
(\Delta +\rho -\epsilon)F_2 +(\delta-\alpha)F_1 &=& i m_* \, G_2, \nonumber\\
(D-\epsilon +\rho)G_2 -(\delta-\alpha)G_1 &=& i m_* \, F_2, \nonumber\\
(\Delta -\rho +\epsilon)G_1 -(\delta^*-\alpha)G_2 &=& i m_* \, F_1\,,
\end{eqnarray}
so that the spin coefficients are involved only through a real $\alpha$ and the combination
\beq
\epsilon-\rho=-\epsilon=\frac{i\sqrt{2}}{4}\Omega\,, 
\eeq
which is purely imaginary.
Hence, Dirac equation can be finally expressed as
\begin{eqnarray}
(D-\epsilon)F_1 +(\delta^*-\alpha)F_2 &=&i m_* \, G_1, \nonumber\\
(\Delta +\epsilon)F_2 +(\delta-\alpha)F_1 &=& i m_* \, G_2, \nonumber\\
(D+\epsilon)G_2 -(\delta-\alpha)G_1 &=& i m_* \, F_2, \nonumber\\
(\Delta -\epsilon)G_1 -(\delta^*-\alpha)G_2 &=& i m_* \, F_1\,.
\end{eqnarray}
Let us look for separable solutions of the form
\beq
\label{Dirac10}
[F_1,F_2,G_1,G_2]=[X_1(x),X_2(x),Y_1(x),Y_2(x)]e^{-i\omega t +ik_2y +ik_3 z}\,\quad
\eeq
with $k_2=0$, as for scalar waves examined in the previous section.
 
The resulting equations for the spinor components $X_{1,2}$ and $Y_{1,2}$ form a first order system of ordinary differential equations with constant (complex) coefficients, namely,  
\beq
\label{sys_dirac}
\frac{d}{dx} \,
\left(
\begin{array}{c}
X_1\cr
X_2\cr
Y_1\cr
Y_2
\end{array}
\right) = 
\left(
\begin{array}{cccc}
A_+& B_+& 0& b \cr
B_-& A_-& b & 0\cr
0 & -b & A_+& C_+\cr
-b & 0 & C_-& A_-
\end{array}
\right)\,
\left(
\begin{array}{c}
X_1\cr
X_2\cr
Y_1\cr
Y_2
\end{array}
\right) \,,
\eeq
where $b=i {\hat m}_0$ and
\beq
A_\pm = -\frac{\lambda}{2}\pm  \omega \eta  \,,\quad
B_\pm = i \left[ -\omega \pm\left(k_3+\frac{\Omega}{2} \right)   \right]\,,\quad
C_\pm = i \left[\omega   \pm    \left(k_3 -\frac{\Omega}{2} \right)\right]\,.
\eeq
Let us note that the function ${\mathcal U}= \exp(\lambda x)$ appears in Eq.~\eqref{Dirac1} and the subsequent analysis in association with variation along the $y$ direction, but such variation disappears once we set $k_2=0$ in Eq.~\eqref{Dirac10}.

The eigenvalues of the $4\times 4$ matrix in Eq.~\eqref{sys_dirac} determine the solutions. These are given by
\beq
\Lambda_{1,\ldots , 4}=-\frac{\lambda}{2}+\Sigma \Big[\omega^2(\eta^2-1)+\left(\sqrt{{\hat m}_0^2+k_3^2}+\Sigma' \frac{\Omega}{2}\right)^2\Big]^{1/2}\,,
\eeq
where $\Sigma$ and $\Sigma'$ independently can be either $+1$ or $-1$; that is, $\Sigma^2=\Sigma'^2=1$. The elementary solutions of system~\eqref{sys_dirac} are of the form $e^{\Lambda_I x}$, where $I=1,\ldots , 4$. We require that the Dirac equation have
 finite solutions for all the spinor components for $x: - \infty \to \infty$.
The only possibility is then that of having constant solutions for $X_{1,2}$ and $Y_{1,2}$, i.e., the determinant of the $4\times 4$ matrix in Eq.~\eqref{sys_dirac} must vanish. Thus the allowed values of $\omega$ satisfy the relation
\beq
\label{spectrum}
\omega^2(\eta^2-1)+\left(\sqrt{{\hat m}_0^2+k_3^2}\pm \frac{\Omega}{2}\right)^2-\frac{\lambda^2}{4}=0\,,
\eeq
which agrees with Eq.~\eqref{g-t9} in the WKB regime. That is, with $E=\hbar \omega$ and $P_3=\hbar k_3$, Eq.~\eqref{spectrum} can be written as 
\beq
\label{spectrum'}
(\eta^2-1)E^2+P_3^2+m_0^2\pm \hbar \Omega (P_3^2+m_0^2)^{1/2}+\frac{1}{4}\hbar^2 \Omega^2 (1- 4 \eta^{-2})=0\,,
\eeq
so that as $\hbar \to 0$, while $E$, $P_3$ and $m_0$ are held fixed, we recover the WKB limit. 

The two branches of Eq.~\eqref{spectrum'} reflect the spin $1/2$ character of the Dirac particle; indeed, with $m_0=0$, the helicity-rotation coupling evident in the spectrum~\eqref{spectrum'} is precisely reminiscent of the corresponding result for photons~\cite{bm}, when one allows for the different helicities involved. 

For a localized packet of cosmic ray protons, for instance, we expect that the packet is dominated by $k_3>>\lambda$ and $k_3>>\Omega$, where $1/\lambda$ and $1/\Omega$ are length scales associated with the G\"odel-type universe. Thus propagation of the packet parallel to the rotation axis is generally possible in the causal case ($\eta<1$) and impossible in the acausal case ($\eta>1$). In the latter situation, we expect confinement of the \emph{high-energy} particles in space, just as for the scalar field---see Eqs.~\eqref{g-t9''} and~\eqref{g-t9'''}---since waves with  $k_3<\lambda$ and $k_3<\Omega$ may be able to propagate in accordance with Eq.~\eqref{spectrum}.

\section{Discussion}

The most basic measurements of an observer consist of the determination of temporal and spatial intervals. In this connection, we have explored the admissibility of coordinate patches in some open spacetime region on the basis of splitting spacetime into its elements via the threading and slicing approaches. All coordinate systems in general relativity must satisfy the threading (Landau-Lifshitz) conditions. The Lichnerowicz admissibility conditions involve the additional slicing coordinate requirements, which protect chronology and exclude CTCs. 

If it were possible to incorporate the slicing coordinate conditions into the geometric structure of Lorentzian geometry, violations of chronology and hence CTCs would be forbidden in the general theory of relativity. However, since the ``local" $t=$ constant hypersurfaces cannot be naturally absorbed into the foundations of general relativity, the slicing conditions simply place additional restrictions on possible coordinate systems. 

The threading and slicing coordinate requirements together ensure that the coordinates assigned to spacetime events are physically reasonable. As such, coordinate systems that can be physically constructed, such as the GPS system, are expected to satisfy Lichnerowicz coordinate conditions. We have examined the physical implications of Lichnerowicz conditions by studying wave propagation in spacetimes that do not possess universal temporal coordinates. In such stationary rotating G\"odel-type universes, for instance, we show that radiation in the WKB approximation cannot propagate parallel to the axis of rotation. Possible generalizations of this result as well as further consequences of chronology violation for wave propagation are topics for future investigation.

\begin{acknowledgments}

Helpful discussions with Robert Jantzen, Sergei Kopeikin and Yuri Obukhov are gratefully acknowledged.

\end{acknowledgments}

\appendix

\section{General Form of Equation~\eqref{A6}}

Imagine an arbitrary $(1+n)\times (1+n)$ matrix of the form
\beq
A=\left(\begin{array}{cccc}
A_{00}& A_{01} &\ldots & A_{0n}\\
A_{10}& A_{11} &\ldots & A_{1n}\\ 
\vdots & \vdots   &\ldots & \vdots  \\
A_{n0}& A_{n1} &\ldots & A_{nn}\\
\end{array}
 \right)\,,
\eeq
which can be denoted as
\beq
A=\left(
\begin{array}{cc}
A_{00}& A_{0i} \\
A_{i0}& A_{ij}\\
\end{array}
 \right)\,,\qquad i,j=1,2,\ldots, n\,.
\eeq
We assume that $A_{00} \ne 0$; in fact, this condition can be ensured---unless $A$ is trivial in the sense that it has only zeros  in its first row and column---by appropriately shifting rows or columns, as we are only interested in ${\rm det} (A)$. Then, it is possible to show that
\beq
\label{eq:A3}
{\rm det} (A)=A_{00}~{\rm det}\left(A_{ij}-\frac{A_{i0}A_{0j}}{A_{00}}\right)\,.
\eeq
Consider the first row of matrix $A$,
\beq
R_0=[A_{00},A_{01},\ldots , A_{0n}]\,.
\eeq
Now subtract $R_0 (A_{10}/A_{00})$ from the second row of matrix $A$, $R_0 (A_{20}/A_{00})$ from the third row and so on.
In this process, the determinant of the matrix does not change due to its alternating character but matrix $A$ is transformed into
\beq
B=\left(
\begin{array}{cc}
A_{00}& A_{0i} \\
0& A_{ij}-\frac{A_{i0}A_{0j}}{A_{00}}\\
\end{array}
 \right)\,.
\eeq
The determinant of this matrix is simply
\beq
{\rm det} (B)=A_{00}~{\rm det}\left(A_{ij}-\frac{A_{i0}A_{0j}}{A_{00}}\right)\,,
\eeq
which proves Eq. (\ref{eq:A3}), since ${\rm det} (A)={\rm det} (B)$.

\end{document}